\documentstyle[12pt,epsf]{article}
\def\fun#1#2{\lower3.6pt\vbox{\baselineskip0pt\lineskip.9pt
\ialign{$\mathsurround=0pt#1\hfil##\hfil$\crcr#2\crcr\sim\crcr}}}

\newcommand{\be}{\begin{equation}}
\newcommand{\ee}{\end{equation}}
\newcommand{\bd}{\begin{displaymath}}
\newcommand{\ed}{\end{displaymath}}
\newcommand{\ba}{\begin{array}}
\newcommand{\ea}{\end{array}}
\newcommand{\bt}{\begin{tabular}}
\newcommand{\et}{\end{tabular}}

\begin{document}

\begin{center}
\large
ELECTROMAGNETIC FORM FACTORS OF THE TRANSITION 
 $\pi+\gamma^{\ast} \rightarrow A_1$.
\end{center}

\begin{center} 
{\bf I.A. Shushpanov}
\end{center}

\begin{center}
\it{Institute for Theoretical and Experimental Physics, 
B. Cheremushkinskaya 25, Moscow 117259, Russia}
\end{center}

\vspace{1.cm}
\centerline{December 1996}
\vspace{1.5 cm}

\begin{abstract}

 Corrections due to gluon condensate to electromagnetic
 form factors of the transition
$\pi+\gamma^{\ast} \rightarrow A_1$ are calculated by standard QCD sum rules
 technique[1,2]. The obtained results are compared to the corresponding
 light-cone
  QCD sum rules[3].
\end{abstract}

\section{}
 The way of deriving sum rules in the framework of QCD is well-known.
 To obtain sum rules for electromagnetic form factors of
 $\pi+\gamma^{\ast} \rightarrow A_1$ transition one should consider
 the three-point function:

\be
\Gamma_{\mu\nu;\lambda}(p,p';q)=-\int dxdye^{ip'x-iqy}
<0|T\left\{j^+_{A\nu}(x)j^{el}_{\lambda}(y)j_{A\mu}(0)\right\}|0>
\ee
where $j_{A \mu}$ is the axial vector light quark current,
$j_{A\mu}=\bar u(x)\gamma_{\mu}\gamma_5d(x);$
$j^{el}_{\lambda}(x)=e_u\bar u(x)\gamma_{\mu}u(x) +e_d\bar d(x)\gamma_{\mu}d(x)$
is the electromagnetic current and $q=p'-p.$

 In the euclidean region $p^2$, $p'^2$, $q^2<0$ ,
  when the virtualities are large enough,
 the amplitude can be obtained in the framework of pertubative QCD,
 the corresponding triangle graphs are represented in Fig. 1.
 If virtualities are not so large then correction to the lowest
 order appears. There are two types of these corrections:
 pertubative corrections and non-pertubative ones arising from
 non-trivial interactions with vacuum fields. In the following
 the virtuality region
  $|p^2| \sim |p'^2|\sim |q^2|\sim 1$ GeV$^2$ will be important,
   where $\alpha_{s} / \pi \sim0.1$  , pertubative
 corrections are small and can be disregarded.

 On the contrary,  non-pertubative corrections which decrease
 as a power law are very important here. The systematic treatment
 of these corrections can be carried out in the operator product
 expansion. It is clear from dimensional considerations that
 the corrections due to gluonic $<0|G^a_{\mu\nu}G^a_{\mu\nu}|0>$
 and quark condensates $<0|\bar\psi\psi|0>^2$  give main
 contributions.

  The amplitude has a complicate tensor structure and it is convenient
  to represent it in the form:
$\Gamma_{\mu\nu;\lambda}(p,p';q)=\sum{f_i(p,p';q)e^i_{\mu\nu;\lambda}}$
  where $e^{i}_{\mu\nu ;\lambda}$ are all structures which can be
  built from $p_{\alpha},$ ${p'}_{\alpha},$ $g_{\alpha\beta}$ and $f_i(p,p';q)$
  are the corresponding coefficients.

    The following structures:
 $p'_{\mu}g_{\nu\lambda}+p_{\nu}g_{\mu\lambda},$
 $p_{\mu}g_{\nu\lambda}+p'_{\nu}g_{\mu\lambda},$
 $q_{\lambda}(p_{\mu}p_{\nu}-p'_{\mu}p'_{\nu})$ will be used
 to determine  form factors.

 Let us consider the coefficient of some structure and write down the
 dispersion relation over axial currents virtualities $p^2,$ $p'^2,$
 holding $Q^2=-q^2$ fixed:
\be
f_i(p^2,p'^2;q^2)=\int\limits^{\infty}_0ds\int\limits^{\infty}_0ds'
\frac{\rho_i(s,s',q^2)}{(s-p^2)(s'-p'^2)}+\mbox{subtr. terms}
\ee

  Quantity $\rho_i(p^2,p'^2,q^2)$
 equals the double discontinuity of the amplitude $f_i(p^2,p'^2,q^2)$
 on the cuts $0\le p^2,p'^2\le\infty$
 divided by $-4\pi^2.$ To get rid of the unknown subtraction terms,
 which are polynomial in one of variables $p^2$ or $p'^2$ but arbitrary
 functions in the other two, it is convenient to use the Borel
 transformation in variables $p^2$ and $p'^2$ simultaneously:
\be
B_{M^2}B'_{M^2}f_i(p^2,p'^2,Q^2)=\int\limits^{\infty}_0ds
\int\limits^{\infty}_0ds'e^{-\frac{s+s'}{M^2}}\rho_i(s,s',Q^2)
\ee

 Parameters $M^2$ and $M'^2$ are chosen equal. The value of $M^2$
 should be optimized so the power corrections and contribution of
 higher states are small (say less than $30 \%$).

 Spectral function $\rho_{\mu\nu;\lambda}(p^2,p'^2,Q^2)$
 can be represented as a sum over physical
 hadronic states. The value of $M^2$ will turn out to be 1 GeV$^2$
 and it means that the main contributions to Borel-transformed
 amplitude will give $\pi$ and $A_1$ mesons while the contributions
 of higher states will be exponentially suppressed. The contributions
 of $\pi$ and $A_1$ mesons will be explicitly accounted for 
 and for the contribution of continuum the following model of
 spectral function will be adopted[2]:

\be
\rho^{cont}_{\mu\nu;\lambda}(p^2,p'^2,Q^2)
=[1-\theta(s_0-s-s')]\rho^{quark}_{\mu\nu;\lambda}(p^2,p'^2,Q^2)
\ee
 where $\rho^{quark}$ is the spectral function calculated in
 the pertubative QCD, $s=p^2$, $s'=p'^2$ and $s_0$ is the continuum threshold.

\section{}
 To obtain explicitly the sum rules for the electromagnetic form factors
 one should match Borel-transformed phenomenological part,
 consisting of the contributions of $\pi$ and $A_1$ mesons,
 and theoretical part, consisting of unity operator (i.e. pertubative
 part), gluon $<0|G^a_{\mu\nu}G^a_{\mu\nu}|0>$ and fermion
 $<0|\bar\psi\psi|0>^2$ condensates.

 Using the following formulae:
\be
<0|j_{A_{\mu}}(0)|\pi^+(p)>=ip_{\mu}f_{\pi}
\ee

\be
<0|j_{A_{\mu}}(0)|A^+_1(p,\varepsilon)>=\sqrt2\frac{m^2_A}{g_A}
\varepsilon_{\mu}
\ee

\be
\ba{ll}
<A^+_1(p',\varepsilon')|j^{el}_{\lambda}(0)|A^+_1(p,\varepsilon)>=
-\varepsilon'_{\sigma}\varepsilon_{\rho}\left\{\left[
{\cal P}_{\lambda}g_{\rho\sigma}-p'_{\rho}g_{\lambda\sigma}-
p_{\sigma}g_{\lambda\rho}\right]F^A_1(Q^2)+\right.\\[4mm]
\left.+\left[g_{\lambda\rho}q_{\sigma}-g_{\lambda\sigma}q_{\rho}\right]
F^A_2(Q^2)+\frac{1}{m^2_A}p'_{\rho}p_{\sigma}{\cal P}_{\lambda}
F^A_3(Q^2)\right\}
\ea
\ee

\be
\ba{ll}
<\pi^+(p')|j^{el}_{\lambda}(0)|A^+_1(p,\varepsilon)>=-\frac{1}{m_A}
\left\{\left[({\cal P}q)g_{\lambda\sigma}-{\cal P}_{\lambda}
q_{\sigma}\right]G_1(Q^2)+\right.\\[4mm]
\left.+\frac{1}{m^2_A}\left[({\cal P} q)q_{\lambda}-q^2{\cal P}_{\lambda}
\right]p'_{\sigma}G_2(Q^2)\right\}\varepsilon_{\sigma};\quad
{\cal P}_{\lambda}=p'_{\lambda}+p_{\lambda};\quad
q'_{\lambda}=p'_{\lambda}-p_{\lambda}
\ea
\ee

 one can explicitly obtain the contributions of $\pi$ and $A_1$ mesons
 in the structures under study:

$p'_{\mu}g_{\nu\lambda}+p_{\nu}g_{\mu\lambda}$:

\be
\ba{lll}
B_{M^2}B'_{M^2}f_{1}(p^2,p'^2;Q^2)[\pi\gamma\to\pi]=0\\[4mm]
B_{M^2}B'_{M^2}f_{1}(p^2,p'^2;Q^2)[\pi\gamma\to
A_1]=0\\[4mm]
B_{M^2}B'_{M^2}f_{1}(p^2,p'^2;Q^2)[A_1\gamma\to
A_1]=\frac{2m^4_A}{g^2_A}e^{-\frac{2m^2_A}{M^2}}
\left(F^A_1(Q^2)+F^A_2(Q^2)\right)
\ea
\ee

$p_{\mu}g_{\nu\lambda}+p'_{\nu}g_{\mu\lambda}$:

\be
\ba{lll}
B_{M^2}B'_{M^2}f_{2}(p^2,p'^2;Q^2)[\pi\gamma\to\pi]=0\\[4mm]
B_{M^2}B'_{M^2}f_{2}(p^2,p'^2;Q^2)[\pi\gamma\to
A_1]=-f_{\pi}\sqrt2\frac{m^3_A}{g_A}e^{-\frac{m^2_A}{M^2}}G_1(Q^2)\\[4mm]
B_{M^2}B'_{M^2}f_{2}(p^2,p'^2;Q^2)[A_1\gamma\to
A_1]=-\frac{2m^4_A}{g^2_A}\left(1+\frac{Q^2}{2m^2_A}\right)
e^{-\frac{2m^2_A}{M^2}}\left(F^A_1(Q^2)+F^A_2(Q^2)\right)
\ea
\ee

$q_{\lambda}(p_{\mu}p_{\nu}-p'_{\mu}p'_{\nu})$:

\be
\ba{lll}
B_{M^2}B'_{M^2}f_{3}(p^2,p'^2;Q^2)[\pi\gamma\to\pi]=0\\[4mm]
B_{M^2}B'_{M^2}f_{3}(p^2,p'^2;Q^2)[\pi\gamma\to
A_1]=f_{\pi}\sqrt2\frac{m_A}{g_A}e^{-\frac{m^2_A}{M^2}}G_2(Q^2)\\[4mm]
B_{M^2}B'_{M^2}f_{3}(p^2,p'^2;Q^2)[A_1\gamma\to
A_1]=\frac{m^2_A}{g_A}e^{-\frac{2m^2_A}{M^2}}
\left(F^A_1(Q^2)+F^A_2(Q^2)\right)
\ea
\ee

 The contribution of unity operator and
 fermion condensate was calculated in [2]:

$p'_{\mu}g_{\nu\lambda}+p_{\nu}g_{\mu\lambda}$:

\be
\ba{ll}
B_{M^2}B'_{M^2}f_{1}(p^2,p'^2;Q^2)[I]=\frac{M^4}{32\pi^2}
\int\limits^{\infty}_0dxe^{-x}\left\{-3k\ln\left(1+\frac{2x}{k}\right)
+\frac{x(16x^2+18xk+6k^2)}{(2x+k)^2}\right\}\\[4mm]
B_{M^2}B'_{M^2}f_{1}(p^2,p'^2;Q^2)[<\bar\psi\psi>]=\frac{4\pi\alpha_s}{81}
<\bar\psi\psi>^2\frac{1}{M^2}[-20-8k]
\ea
\ee

\vspace{10mm}

$p_{\mu}g_{\nu\lambda}+p'_{\nu}g_{\mu\lambda}$:

\be
\ba{ll}
B_{M^2}B'_{M^2}f_{2}(p^2,p'^2;Q^2)[I]=-\frac{M^4}{8\pi^2}
\int\limits^{\infty}_0dxe^{-x}\frac{5x+3k}{(2x+k)^2}x^2\\[4mm]
B_{M^2}B'_{M^2}f_{2}(p^2,p'^2;Q^2)[<\bar\psi\psi>]=\frac{4\pi\alpha_s}{81}
<\bar\psi\psi>^2\frac{1}{M^2}[64+5k-2k^2]
\ea
\ee

$q_{\lambda}(p_{\mu}p_{\nu}-p'_{\mu}p'_{\nu})$:

\be
\ba{ll}
B_{M^2}B'_{M^2}f_{3}(p^2,p'^2;Q^2)[I]=\frac{3M^2}{16\pi^2}
\int\limits^{\infty}_0dxe^{-x}\left\{\ln\left(1+\frac{2x}{k}\right)-
\frac{2x}{2x+k}\right\}\\[4mm]
B_{M^2}B'_{M^2}f_{3}(p^2,p'^2;Q^2)[<\bar\psi\psi>]=\frac{4\pi\alpha_s}{81}
<\bar\psi\psi>^2\frac{1}{M^4}\left[5-\frac{18}{k}+\frac{18}{k^2}+
2k\right],
\ea
\ee
where $k=\frac{Q^2}{M^2}$.

 The corresponding sum rules, which include only these operators,
 are incomplete because gluon condensate has dimension even less
 that fermion one and its contribution also should be accounted for.

\section{}

  The contribution proportional $<G^2>$ into amplitude
  $\Gamma_{\mu\nu;\lambda}(p^2,p'^2;Q^2)$ could be obtained after
  taking into account the interaction of quarks, running through
  triangle loop in Fig. 1, with external vacuum gluon field.
  In the first order of $\alpha_s$ there are 12 diagrams and 6 of them
  are represented in Fig. 2. The other 6 diagrams correspond to the case when
  electromagnetic current enters the d-quark line.

  In fixed-point gauge $x_{\mu}A^a_{\mu}(x)=0$  the potential
  $A^a_{\mu}(x)$ is expressed through the field strength and its
  covariant derivatives at the origin 
(black circle in the left lower angle):

$A^a_{\mu}(x)=-\frac12x_{\nu}G^a_{\mu\nu}(0)-\frac13x_{\nu}x_{\alpha}
(D_{\alpha}G_{\mu\nu})^a(0)+\ldots$

  Graphs in which there is only one external gluon line disappear
  after averaging over vacuum gluonic fields, because v.e.v of
  $G^a_{\mu \nu}$, $D_{\lambda}G^a_{\mu \nu}$ equals zero.
  Non-zero contributions appear from the graphs with two
  external gluon lines and vacuum averaging is performed according
  to the rule:

\be
<0|G^a_{\mu\nu}G^b_{\lambda\rho}|0>=\frac{1}{96}\delta^{ab}
<0|G^a_{\mu\nu}G^a_{\mu\nu}|0>(g_{\mu\lambda}g_{\nu\rho}-
g_{\mu\rho}g_{\nu\lambda})
\ee

   In fixed-point gauge the diagrams in Figs.2 e,f are equal to zero[4].
Expressions for contributions of the other diagrams in
 $\Gamma_{\mu\nu;\lambda}(p^2,p'^2;Q^2)$ can be obtained
 if one consider
 the equation for the quark propagator in the external gluonic field:

\be
i\gamma_{\mu}\left(\partial_{\mu}+ig\frac{\lambda^a}{2}A^a_{\mu}(x)\right)
\cdot S(x,z)=i\delta^4(x-z)
\ee

 Solving this equation up to the second order of $A$ we get:

\be
\ba{ll}
S(x,z)=i\int\frac{d^4k}{(2\pi)^4}e^{-ik(x-z)}\left\{\frac{\hat k}{k^2}
-\frac14g\lambda^aG^a_{\alpha\beta}\cdot\varepsilon_{\alpha\beta\sigma\rho}
\gamma_5\gamma_{\rho}\frac{k_{\sigma}}{k^4}+\right.\\[4mm]
\left.+\frac14g\lambda^aG^a_{\alpha\beta}z_{\beta}(\gamma_{\alpha}k^2-
2k_{\alpha}\hat k)\cdot\frac{1}{k^4}+\frac{1}{96}g^2
<0|G^2_{\mu\nu}|0>\cdot\left[-z^2\frac{\hat k}{k^4}-
4\frac{\hat k}{k^6}(kz)^2+2\hat z\frac{kz}{k^4}\right]\right\}
\ea
\ee

  The power corrections proportional to $<0|G^a_{\mu\nu}G^a_{\mu\nu}|0>$
  can be calculated if one substitutes propagator $S(x,z)$ (17)
  instead of each fermionic line in Fig. 1, selects the terms proportional
  to $<G^2>$ averages over vacuum fluctuations of the gluonic field
  according to (15) and goes to the momentum representation.
  As a result, for amplitude with gluon emission from different lines
  we have:

\be
\ba{llll}
\Gamma^{Fig.2(a+b+c)}_{\mu\nu;\lambda}(p^2,p'^2;Q^2)=\frac{ig^2}{24}
<0|G^2_{\mu\nu}|0>\int\frac{d^4k}{(2\pi)^4}
\frac{1}{k^2(p-k)^2(p'-k)^2}\times\\[4mm]
\times\left(Sp\left\{\hat k\gamma_{\nu}[2\gamma_{\lambda}(p-k,p'-k)+
(\hat p-\hat k)\gamma_{\lambda}(\hat p'-\hat k)]\gamma_{\mu}\right\}
\frac{1}{(p-k)^2(p'-k)^2}+\right.\\[4mm]
+Sp\left\{[2\gamma_{\nu}(\hat p'-\hat k)\gamma_{\lambda}(pk-k^2)+
(\hat p-\hat k)\gamma_{\nu}(\hat p'-\hat k)\gamma_{\lambda}\hat k]
\gamma_{\mu}\right\}\frac{1}{k^2(p-k)^2}+\\[4mm]
\left.+Sp\left\{[2\gamma_{\nu}(p'k-k^2)+(\hat p'-\hat k)\gamma_{\nu}
\hat k]\gamma_{\lambda}(\hat p-\hat k)\gamma_{\mu}\right\}
\frac{1}{k^2(p'-k)^2}\right)
\ea
\ee

  The contribution of these diagrams structures of interest
  are the following:

$p'_{\mu}g_{\nu\lambda}+p_{\nu}g_{\mu\lambda}$:

\be
B_{M^2}B'_{M^2}f_1(p^2,p'^2;Q^2)[<G^2>]=\frac{g^2<0|G^2_{\mu\nu}|0>}{96\pi^2}
\int\limits^{\infty}_0dxe^{-x}\frac{3x^2-xk-2k^2}{(x+k)^2(2x+k)^2}x
\ee

$p_{\mu}g_{\nu\lambda}+p'_{\nu}g_{\mu\lambda}$:

\be
B_{M^2}B'_{M^2}f_2(p^2,p'^2;Q^2)[<G^2>]=-\frac{g^2<0|G^2_{\mu\nu}|0>}{48\pi^2}
\int\limits^{\infty}_0dxe^{-x}\frac{2k^3+4k^2x+2x^2k+x^3}{(x+k)^2(2x+k)^2}
\cdot\frac xk
\ee

$q_{\lambda}(p_{\mu}p_{\nu}-p'_{\nu}p'_{\mu})$:

\be
B_{M^2}B'_{M^2}f_3(p^2,p'^2;Q^2)[<G^2>]=\frac{g^2<0|G^2_{\mu\nu}|0>}{96\pi^2}
\frac{1}{M^2}
\int\limits^{\infty}_0dxe^{-x}\frac{4k^3-5k^2x+4x^2k+x^3}{(x+k)^4}
\cdot\frac{x}{k^2}
\ee

 The analytical expression for Fig.2d turns out to be

\be
\ba{ll}
\Gamma^{(d)}_{\mu\nu;\lambda}(p^2,p'^2;Q^2)=\frac{ig^2<0|G^2_{\mu\nu}|0>}{192}
\left[\frac{\partial^4}{\partial p_{\alpha}\partial p'_{\alpha}
\partial p_{\beta}\partial p'_{\beta}}-\frac{\partial^2}{\partial
p_{\alpha}\partial p_{\alpha}} \cdot\frac{\partial^2}{\partial
p'_{\beta}\partial p'_{\beta}}\right]\times\\[4mm]
\times\int\frac{d^4k}{(2\pi)^4} \frac{Sp\left\{\gamma_{\mu}(\hat k-\hat
p)\gamma_{\lambda} (\hat k-\hat p')\gamma_{\nu}\hat
k\right\}}{k^2(k-p)^2(k-p')^2} \ea \ee

 The explicit calculations of the contribution of this diagram to
 the interested structures were performed using program "MAPLE"
 and result is rather surprising one: this diagram exactly cancels the others,
 i.e. gluon condensate gives no contribution to the above
 mentioned structures.

\section{}
 The final sum rules for electromagnetic form factors of
 $\pi+\gamma^*\to A_1$ transition are the following:

\be
\ba{ll}
F^A_1+F^A_2+\sqrt2\frac{f_{\pi}g_Ae^{\frac{m_A^2}{M^2}}}{m_A}G_2=
\frac{g^2_A}{m^2_A}e^{\frac{2m^2_A}{M^2}}
\left[\frac{3M^2}{16\pi^2}\int\limits^{\chi_0}_0e^{-x}dx
\left\{\ln\left(1+\frac{2x}{k}\right)-\frac{2x}{2x+k}\right\}+\right.\\[4mm]
\left.+\frac{4\pi\alpha_s}{81}<\bar\psi\psi>^2\cdot\frac{1}{M^4}
\left[2k+5-\frac{18}{k}+\frac{18}{k^2}\right]\right]
\ea
\ee

\be
\ba{ll}
F^A_1+F^A_2+\frac{\sqrt2f_{\pi}m_Ag_A}{Q^2+2m^2_A}e^{\frac{m^2_A}{M^2}}
G_1=\frac{g^2_A}{m^2_A}\cdot\frac{e^{\frac{2m^2_A}{M^2}}}{Q^2+2m^2_A}
\left[\frac{M^4}{8\pi^2}\int\limits^{\chi_0}_0dxe^{-x}
\frac{x^2(5x+3k)}{(2x+k)^2}-\right.\\[4mm]
\left.-\frac{4\pi\alpha_s}{81}<\bar\psi\psi>^2\frac{1}{M^2}
\left\{64+5k-2k^2\right\}\right]
\ea
\ee

\be
\ba{ll}
F^A_1+F^A_2=\frac{g^2_A}{2m^4_A}e^{\frac{2m^2_A}{M^2}}
\left[\frac{M^4}{32\pi^2}\int\limits^{\chi_0}_0dxe^{-x}
\left\{-3k\ln\left(1+\frac{2x}{k}\right)+
\frac{x(16x^2+18xk+6k^2)}{(2x+k)^2}\right\}+\right.\\[4mm]
\left.+\frac{4\pi\alpha_s}{81}<\bar\psi\psi>^2\frac{1}{M^2}
\left\{-20-8k\right\}\right],
\ea
\ee
where $\chi_0=s_0/M^2$.

 These first and second sum rules coincide with the ones obtained
 in [2] but the third one is different.

 The sum rules for $G_1(Q^2)$ and $G_2(Q^2)$ are stable for
 $0.9$ GeV$^2<M^2<1.2$ GeV$^2$.
  Figs.3,4 show dependence of $G_1(Q^2)$ and $G_2(Q^2)$
  upon $M^2$ at the following values of parameters:
 $\alpha_{s}<0|\bar\psi\psi|0>^2=0.8\times 10^{-4}$ GeV$^6$, $g_a=8.9$,
 $m_A=1.26$ GeV, $s_0=3$~GeV$^2.$
 Parameter $g_A$ was determined in [5] using QCD sum rule.
 The left-hand vertical bars
 mark the $M^2$ value above which the ratio of power corrections to
 the sum of terms does not exceed $30\%$. To the left of right-handed
 vertical bars the contribution of the continuum does not exceed $30\%.$
 Choosing $M^2=1.05$ GeV$^2$ we get acceptable region of $Q^2$ for
 $G_1(Q^2)$ and $G_2(Q^2)$:
 $0.4$ GeV$^2<Q^2<2.2$ GeV$^2$ and 
 $0.7$ GeV$^2<Q^2<3.2$ GeV$^2$, respectively.

 There is another calculation method of form factors based on
 light-cone expansion [6,7,8]
 Obtained resultes can be compared
 to the ones derived from the sum rules on light-cone[3] (Figs. 5,6).
 These sum rules are model-dependent and different lines correspond
 to the different wave functions of pion. Three types of wave pion
 function are used: asymptotical [9], Chernyak-Zhitnitsky [10] and
 Braun-Filyanov [11] ones.
 For $G_1(Q^2)$ there is a
 discrepancy between the two methods but, probably, situation can be
 improved after accounting for the operator of a higher dimension.
 For $G_2(Q^2)$ the agreement is sensible and the best choice of the
 wave function is Braun-Filyanov function.

\vspace{5mm}

 I am grateful to B.L. Ioffe for useful discussions and remarks.

\vspace{5mm}

\centerline{\bf Acknowledgements}
 
This work was supported in part
by the International Science Foundation Grant M9H300.

\newpage
\centerline{\bf Figure Captions}

\bigskip

 Fig.1 - The diagrams corresponding to pertubative QCD in the zero order
of $\alpha_s$.  

\bigskip

Fig.2 - The diagrams correspondig to gluon contribution in the first
order of $\alpha_s$.

\bigskip

Fig.3 - The $M^2$-dependence of $G_1$ at fixed $Q^2=0.5$GeV$^2$ 
(solid line) and $Q^2=1.$GeV$^2$ (dashed line). 

\bigskip

Fig.4 - The $M^2$-dependence of $G_2$ at fixed $Q^2=2.$GeV$^2$ 
(solid line) and $Q^2=2.5$GeV$^2$ (dashed line). 

\bigskip

Fig.5 - The $Q^2$-dependence of obtained $G_1$ in comparision to
$G_1$ corresponding to light-cone sume rules.  

\bigskip

Fig.6 -  The $Q^2$-dependence of obtained $G_2$ in comparision to
$G_1$ corresponding to light-cone sume rules.


\newpage

\begin{thebibliography}{11}

\bibitem{}
  M. Shifman, A. Vainshtein and V. Zakharov,  Nucl. Phys. {\bf B147}(1979)
  385,448
\bibitem{}
 B. Ioffe and A. Smilga,  Nucl. Phys. {\bf B216}(1983) 373
\bibitem{}
  V. Belyaev, Z.Phys. {\bf C65}(1995) 93
\bibitem{}
  A. Smilga,  Yad. Fiz. {\bf 35}(1982) 473
\bibitem{}
  L. Reinders, H. Rubinstein and S. Yazakyi  Nucl. Phys.
 {\bf B196}(1982) 125
\bibitem{}
  V. Braun and I. Halperin, preprint TAUP-2137-94
\bibitem{}
  I.Balitsky, V.Braun and A.Kolesnichenko, Sov.J.Nucl.Phys.
 {\bf 44}(1986) 1028; Nucl.Phys. {\bf B312}(1989) 509
\bibitem{}
P.Ball, V.Braun and H.Dosch, Phys.Rev. {\bf D44}(1991) 3567
\bibitem{} 
 V. Chernyak and A. Zhitnitsky,  JETP Lett. {\bf 25}(1977) 510
    Yad. Fiz. {\bf 31}(1980) 1053
\bibitem{} 
 V. Chernyak and A. Zhitnitsky,  Phys. Rep. {\bf 112}(1984) 173
\bibitem{}
 V. Braun and I. Filyanov,  Z.Phys. {\bf C44}(1989) 157; ibid.
 {\bf C48}(1990) 239

\end{thebibliography}
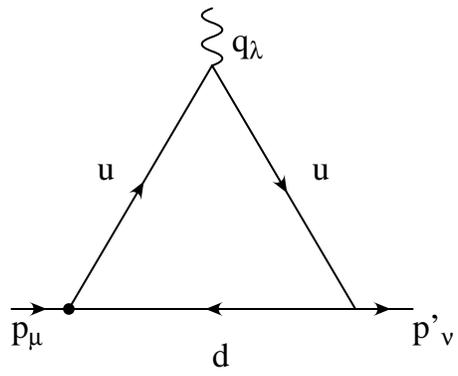
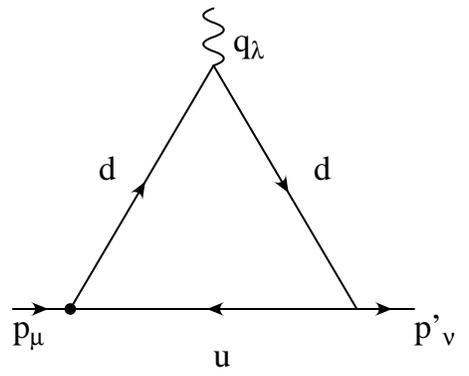
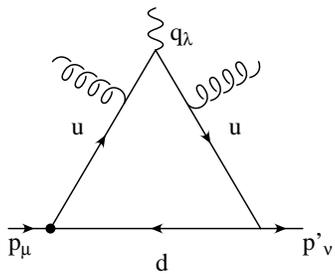
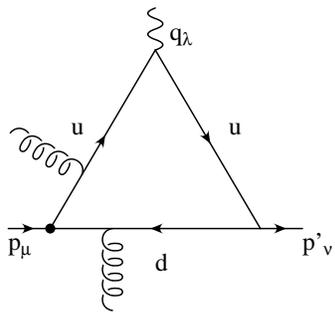
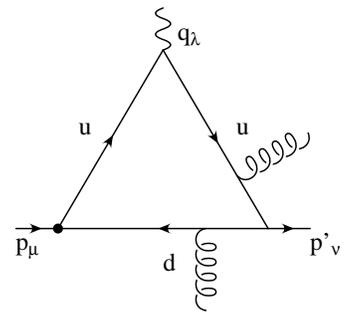
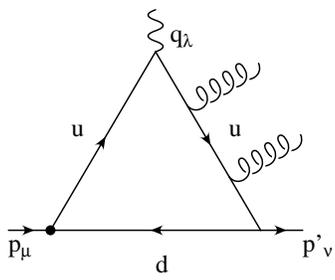
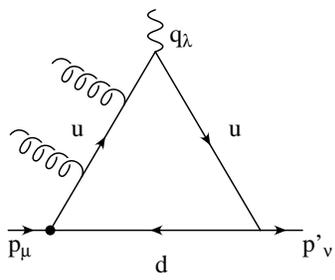
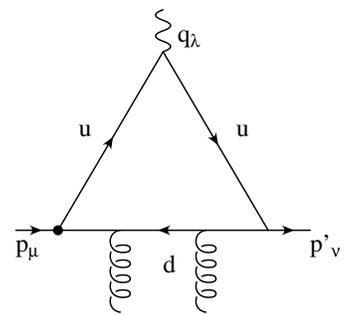
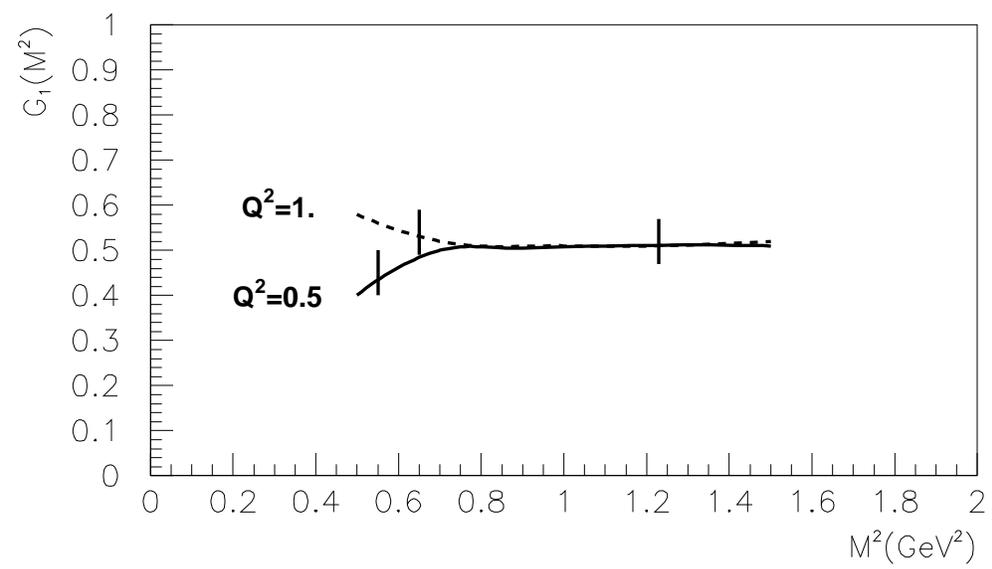
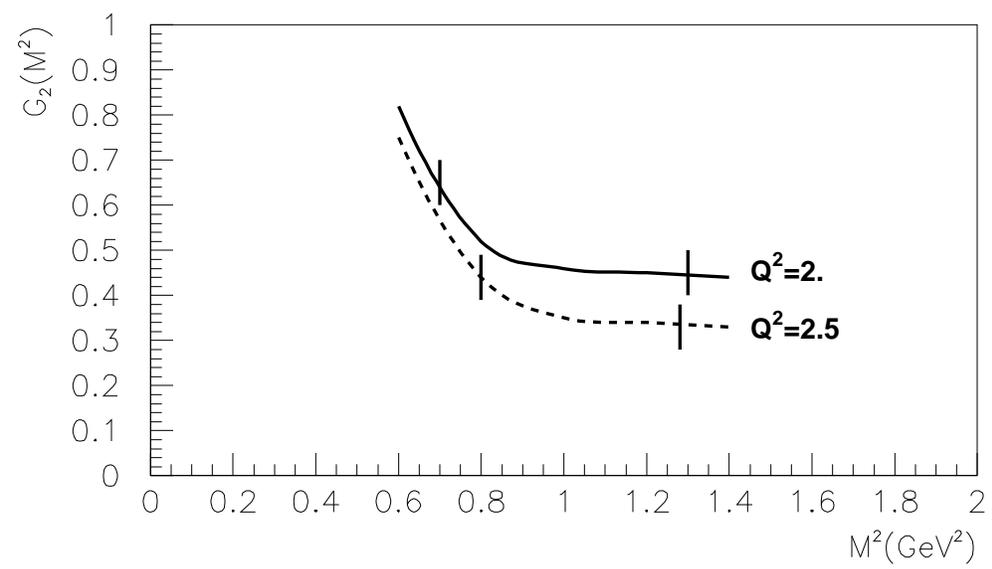
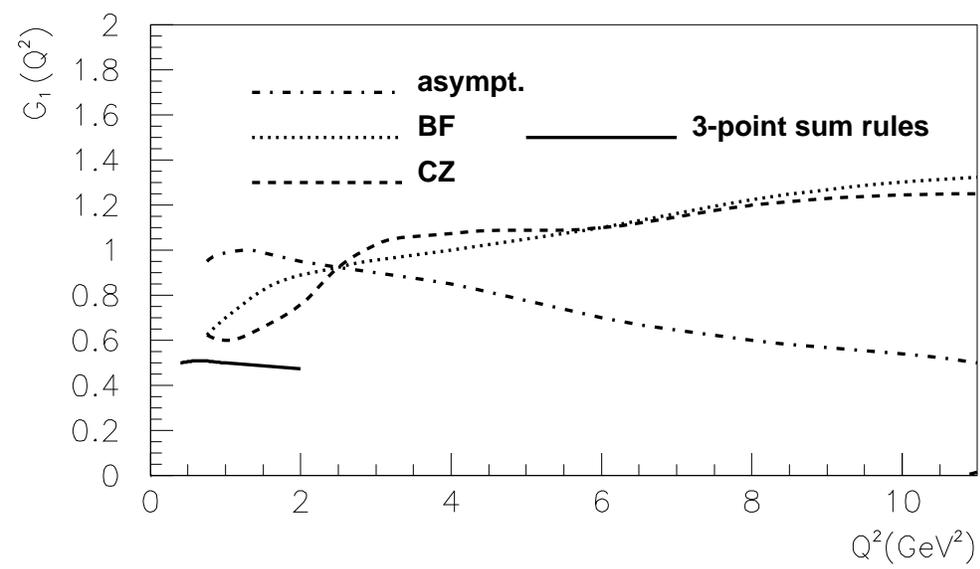
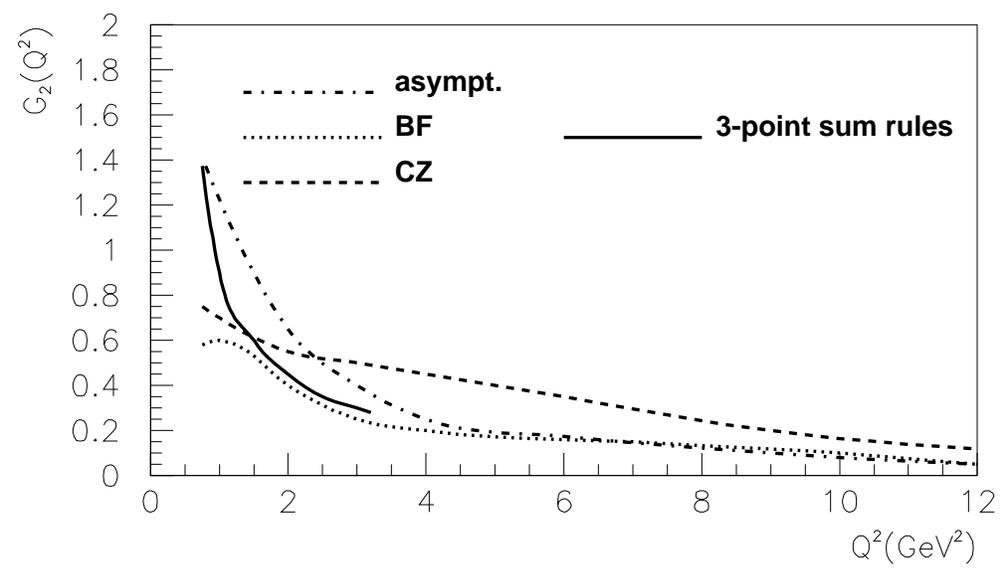
\end{document}